\documentclass{aa}

\input psfig.sty

\font\yfsm=cmb10
\def \NV{N{\yfsm V}}
\def \HeII{He{\yfsm II}}
\def \CIII{C{\yfsm III}]}
\def \CIV{C{\yfsm IV}}
\def \SiIV{Si{\yfsm IV}}
\def \OV{O{\yfsm V}}

\def \NII{N{\yfsm II}}
\def \kiro{km s$^{-1}$}
\def \F10214{IRAS F10214+4724}
\def \SMM{SMM02399-0136}

\def \araa{ARA\&A}
\def \aap{A\&A}
\def \apj{ApJ}
\def \apjl{ApJL}

\def \aj{AJ}
\def \mnras{MNRAS}

\def \nat{Nat}

\begin{document}

\thesaurus{11(11.01.2; 11.17.2; 11.17.4)}

\title{Emission-Line Properties of MG2016+112: 
A Luminous Type-2 Quasar at High Redshift
\thanks{Based on observations with the Canada-France-Hawaii
    Telescope at Mauna Kea, Hawaii, USA.}}

\author{Yamada, T. \inst{1}
\and Yamazaki, S. \inst{1}
\and Hattori, M. \inst{1}
\and Soucail, G. \inst{2} 
\and Kneib, J.-P. \inst{2}}

\offprints{T.Yamada}
\mail{yamada@astr.tohoku.ac.jp}

\institute{Astronomical Institute, Tohoku University, Aoba-ku, 
Sendai 980-77, Japan
\and
OMP, 14 Av. E. Belin, 31400 Toulouse, France}

\date{Received July 20 1999; Accepted}

\maketitle

\begin{abstract}
  We present new high signal-to-noise ratio spectra of the components B
  and C of the gravitational lensing system MG2016+112. We show that
  image C displays strong emission lines of Ly$\alpha$, \NV , \CIV ,
  \HeII , and \CIII\ redshifted to z=3.27, similar to images A and
  B. We examine the emission-line flux ratios in order to put
  constraints on the lens models as well as to investigate the intrinsic
  nature of MG2016+112.  The observed line ratios of B and C
  are consistent with those expected in the simple photo-ionization
  models for narrow-line region of active galactic nuclei (AGN) except
  for the enhanced \NV\ lines. The line ratios difference
  of components B and C can be interpreted as a difference in
  ionization parameters. This result is consistent with lens model
  prediction that C is a fold image of a slightly outer part
  of the nucleus.  MG2016+112 is known to be very unique among the
  high-redshift AGN; it is neither an ordinary broad-line quasar nor
  a powerful radio galaxy as indicated by the width and flux ratio of the 
emission lines. 
 Together with other observed properties discussed in literature,
  we argue that MG2016+112 is the highest redshift luminous
  radio-quiet type-2 quasar.  
\end{abstract}

\section{Introduction}

MG2016+112 is one of the first discovered gravitational-lens
system. Lawrence et al. (1984) observed three distinct radio
sources, A, B, and C. Further higher resolution radio map resolved
C into multiple components (Garrett et al. 1994, 1996). At optical
wavelength, the image A and B are point-source but the image C is
a fainter resolved object (Lawrence et al.  1984; Schneider et
al. 1985, 1986). Spectroscopic observations revealed that A and B show
very similar spectra dominated by strong ultra-violet (UV)
emission-lines redshifted to z=3.27 (Lawrence et al. 1984; Schneider
et al. 1987).  Although the optical spectrum of C has not been fully
published so far
\footnote{Lawrence et al. (1996) show a figure of the spectrum
  obtained with the Keck telescope but there is no detail description.}, 
it has always been considered at the same redshift as A and B after its
detection in the redshifted Ly$\alpha$ narrow-band data 
(Schneider et al. 1986).

The MG2016+112 lens system is enigmatic for 3 reasons. First, the
lensed object itself is very unique among known high-redshift
galaxies (Lawrence et al. 1984). It has been conventionally called a
`quasar' since it is a fairly luminous point-like object. 
However, A and B show only {\it narrow} emission lines and
thus MG2016+112 is not an ordinary broad-line quasar.  Although strong
narrow emission lines are typically seen for high-redshift powerful
radio galaxies (HzPRGs), MG2016+112 appears somewhat different from
the known HzPRGs. Indeed, HzPRGs are typically extended in optical
images and have lobe-dominated radio structures with a scale of a few
tens of kpc while A and B show only point-like features even in 
high-resolution optical and radio images. It is important to
investigate the true nature of the lensed object.

Second, image C is a very complex object: at optical and near-infrared (NIR)
wavelength, C is resolved and has an arc-like morphology.
Radio-to-optical flux ratio of C is several times larger than those of
A and B and cannot be explained by variability and time delay.  At
radio wavelength, C is resolved into two components, C$_1$ and
C$_2$. While A, B and C$_2$ (sometimes referred as C$^\prime$ in
literatures) are point-like objects even at 15-mas resolution, C$_1$
has been further resolved into three chain-like components (Garrett et
al. 1996). Radio spectral shape of C$_2$ is similar to those of A and
B ($\alpha = 0.8$) but C$_1$ has significantly flatter one ($\alpha =
0.2$).  Whether C$_1$ is a radio galaxy at different redshift or a
lensed image of the outer structure of the radio source at z=3.27 is
still in question.
 
Finally the nature of the lens producing this multiple image system is
still an enigma. Deep optical and NIR images have detected a red galaxy
D amid images A, B and C. (Schneider et al.  1986; Langston, Fisher,
\& Aspin 1991; Lawrence, Neugebauer, \& Matthews 1993). Galaxy D
seems to correspond to an evolved giant elliptical galaxy at z$\sim
1$. However, the image separation requires a mass-to-light ratio for D
much larger than typical one. While lens models have assumed the
existence of a high-redshift massive cluster as an additional source of
lensing mass (Narasimha et al. 1984, 1987, 1989; Nair \& Garrett 1997;
Langston et al. 1991), at first no signature of such a cluster has been seen in
the optical and NIR observations (Schneider et al. 1987;
Langston et al. 1991).

Recently, Hattori et al. (1997) detected an extended X-ray emission
in the direction of MG2016+112, possibly emitted by the hot gas in 
the lensing cluster of galaxy. A strong
emission-line-like feature consistent with iron lines
redshifted to z $\sim 1$ was detected. This discovery could in
principle solve the `dark-lens' problem for MG2016+112, but
opened another question of a `dark cluster', namely, the lack of 
optical counterpart of the X-ray hot gas.  Very recently, Kneib et al.
(1997) spectroscopically detected several galaxies at z$\sim 1$ and
Ben\'\i tez et al. (1999) observed a possible color-magnitude sequence
of faint red galaxies in this field arguing for the existence of a distant
massive structure at z$\sim 1$.

We have obtained new spectra of components B, C, D of the MG2016+112
system. In this paper, we analyze the emission lines of
image B and C in order not only to investigate the intrinsic nature
of MG2016+112 but also to understand the component C in view
of the lens-model prediction.  Observations and data
reduction are described in Sect. 2. In Sect. 3, we examine the
observed emission line properties. The line flux ratios are compared with the
prediction of the photo-ionization models as well as those of other
high-redshift objects in various categories. In Sect. 4, the nature of
image C is discussed in the context of lens models. We then argue that
MG2016+112 may be a {\it radio-quiet} type-2 quasar based on
the results obtained in Sect. 3 as well as those discussed in
literatures.
Throughout this paper we use $H_0=50$ km/s/Mpc, $\Omega_0=1$ and $\lambda=0$.

\section{Observations and the Obtained Spectra}

\begin{figure} 
\psfig{file=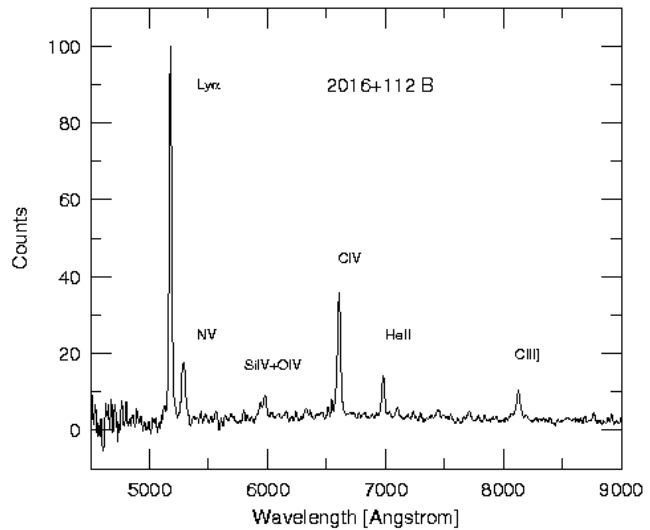,width=0.5\textwidth}
\caption[fig1.ps]{Observed spectra of MG2016+112 B.}  
\end{figure}

Spectroscopic observations were made with the Canada-France-Hawaii
Telescope (CFHT) using the Subarcsecond Imaging Spectrograph (SIS; Le
F\'evre et al. 1994) in August 1997. The Stis2 2048$\times$2048 CCD
with 21 $\mu$m pixel and the R150 grating were used and the resultant
dispersion is 2.88 \AA\ per pixel. The wavelength coverage extends from
4500 to 9000 \AA . Two multi-slit masks were used to obtain
spectra of candidate cluster galaxies in the field. Each slit has
7.8-arcsec-long and 0.78-arcsec-wide aperture, which gives $\sim 18$
\AA\ instrumental resolution.  MG2016+112 B, B$_1$, and D were observed
within one slit of the first mask (Mask1) and C with a slit in the
second one (Mask2). In total, 6 and 5.8 hours exposures were obtained
for Mask1 and Mask2, respectively.

\begin{figure}
\psfig{file=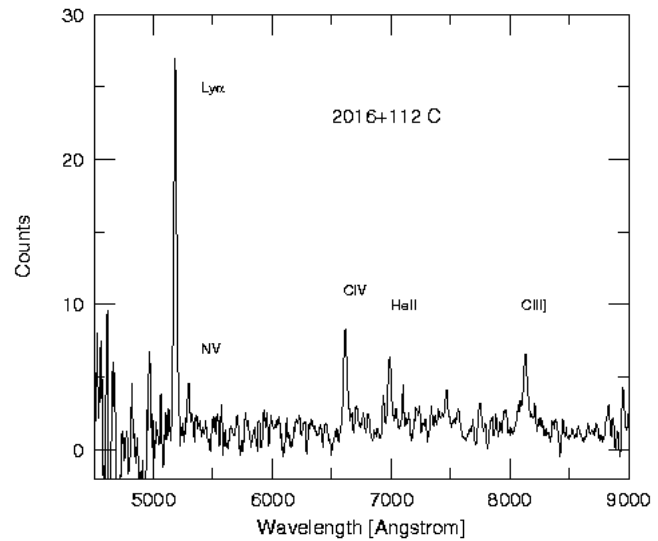,width=0.5\textwidth}
\caption[fig2.ps]{Observed spectra of MG2016+112 C.}  
\end{figure}

The data was pre-reduced using standard IRAF\footnote{IRAF is
distributed by NOAO, which are operated by AURA, Inc., under
cooperative agreement with NSF.}  tasks. We then follow the
reduction procedure of multi-slit data described in Le F\'evre et
al. (1995). Wavelength calibration was done with the arc-line spectra
taken at the observation. Typical internal error of wavelength
determination is $\sim 0.3$ \AA .  The flux calibration was done using
the spectroscopic standard star GD248 (Oke 1990). Note that the flux
calibration becomes more uncertain above 7500-8000\AA\ because
no order-separating filter was used and the flux is contaminated by
the UV and blue contribution of the second order filter. We tried
to carefully take into account this effect but flux calibration may
remain uncertain above 8000 \AA.

Figure~1 and 2 show the obtained spectra of image B and C smoothed
with a 5 pixel gaussian filter. Strong UV emission lines of
Ly$\alpha$, \NV\ $\lambda\lambda 1240$ \AA , \CIV\ $\lambda\lambda
1549$ \AA , \HeII\ $\lambda 1640$ \AA , and \CIII\ $\lambda 1909$ \AA\
are seen in both of the spectra. \SiIV\ and \OV\ lines at 1400 \AA\
are clearly seen in the B spectrum but only marginal in C.
The redshift of C is the same as B within the uncertainties. There
is no sign of contamination by a system at a different redshift. 
 
\section{Emission-Line Properties of MG2016+112 B and C}

\begin{figure} 
\psfig{file=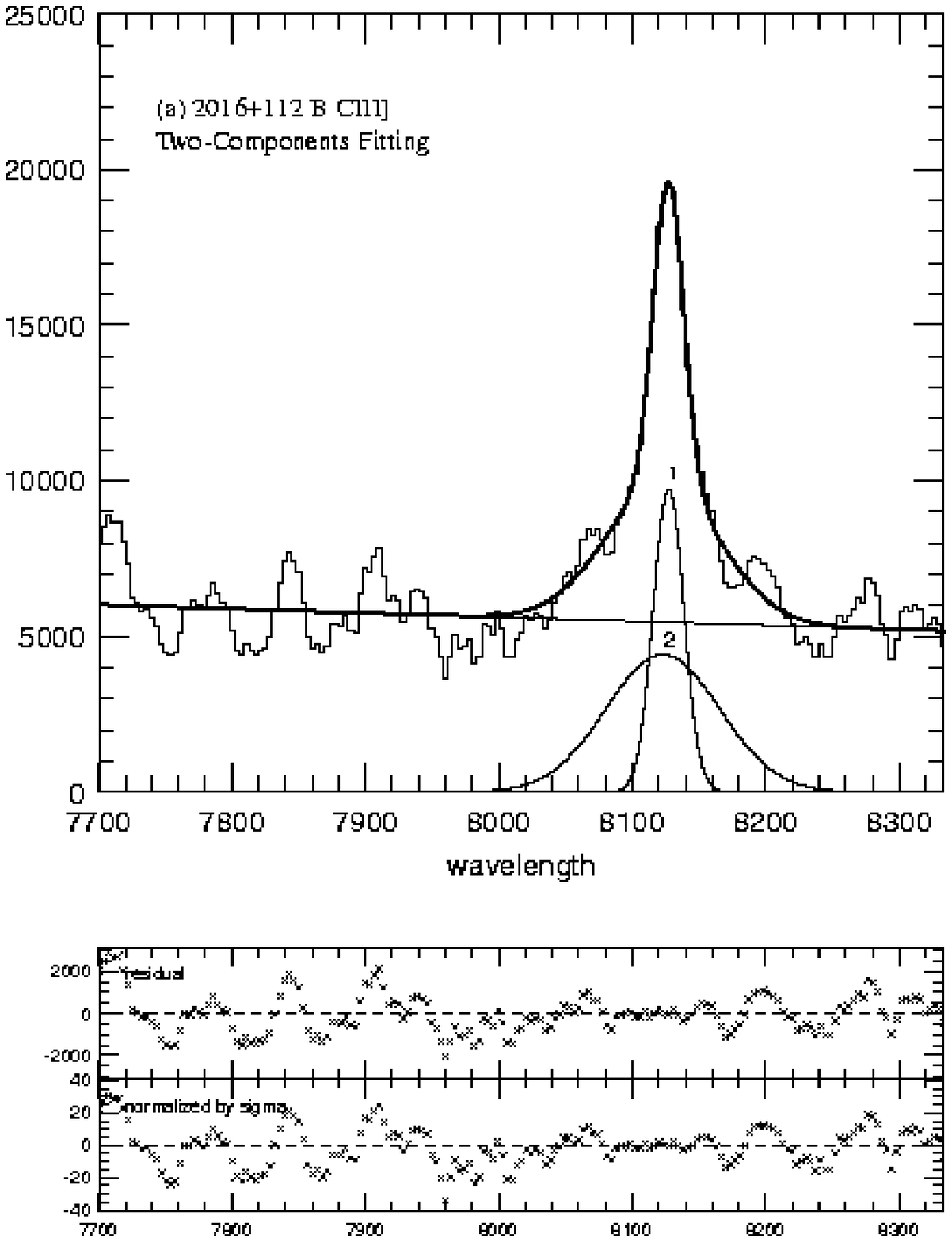,width=0.4\textwidth}
\psfig{file=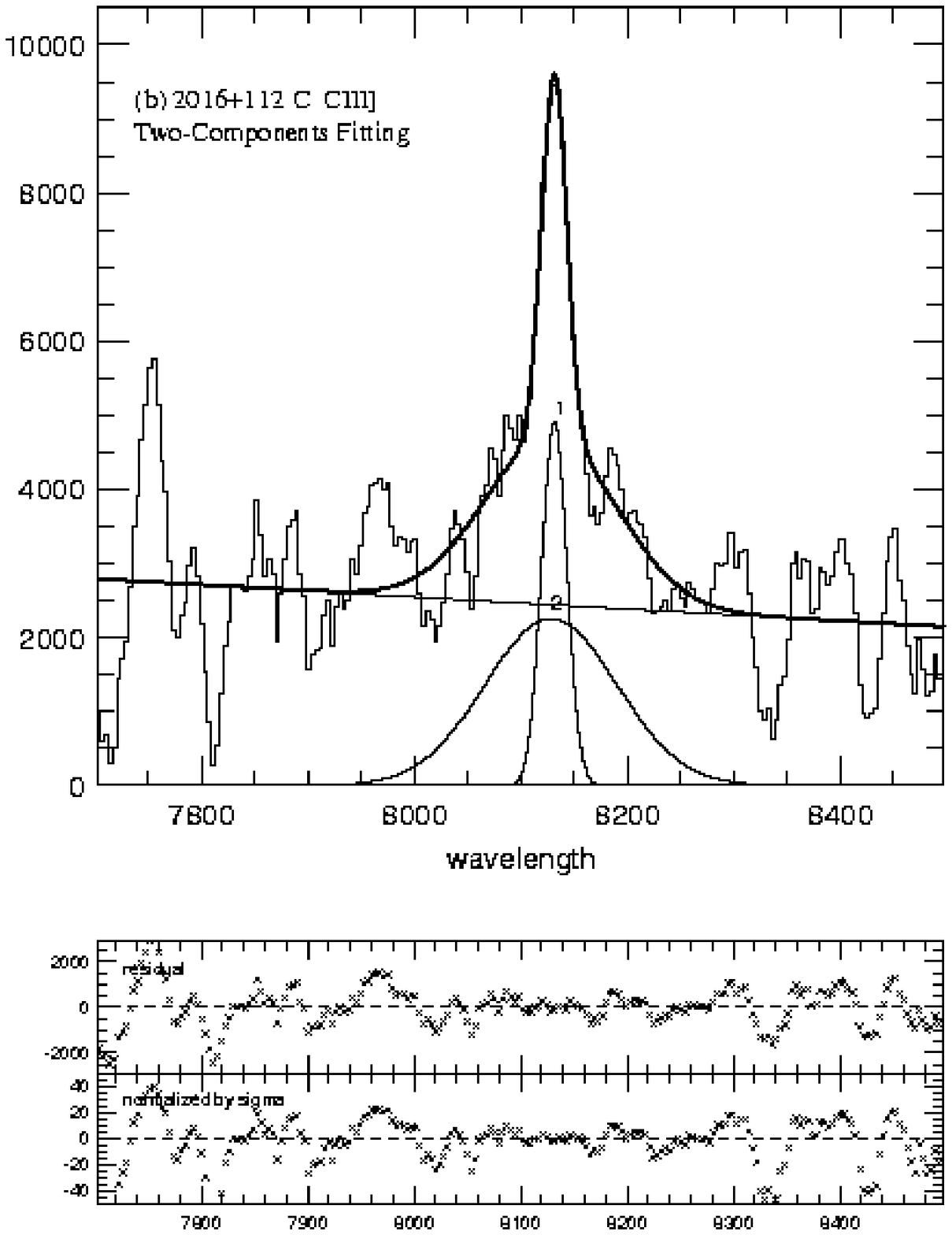,width=0.4\textwidth}
\caption[fig3b.ps]{Results of the two-components Gaussian fitting for
the \CIII\ lines of component B (panel A) and C (panel B).}
\end{figure}

\begin{table*}
\caption{Line width$^{a}$}
\begin{flushleft}
\begin{tabular}{lccccc}
\noalign{\smallskip}
\hline
\noalign{\smallskip}
 & MG2016+112 A$^{b}$ & MG2016+112 B & MG2016+112 C & F10214$^{c}$ & SMM02399$^{d}$ \\
\noalign{\smallskip}
\hline
\noalign{\smallskip}
  Ly$\alpha$      & $<1000$ & $460  $ & $630  $ & $ 900 $ & $1850 $ \\
  \NV\            & $<1600$ & $880  $ & $840  $ & $ 1700$ & $1790 $ \\
  \CIV\           & $<1100$ & $580  $ & $580  $ & $ 1200$ & $1560 $ \\
  \HeII\          & $<1200$ & $460  $ & $440  $ & $ 1150$ & $2800 $ \\
  \CIII\          & ---     & $750  $ & $1100 $ & $     $ & $7700 $ \\
  \CIII\ narrow   & ---     & $360  $ & $380  $ & $ 1000$ & $     $ \\
  \CIII\ broad    & ---     & $1750 $ & $2550 $ & $ 3700$ & $     $ \\
\noalign{\smallskip}
\hline
\noalign{\smallskip}
\end{tabular}
\\
(a) FWHM in km s$^{-1}$ \\
(b) From Lawrence et al. (1984) \\
(c) From Searjeant et al. (1998) \\
(d) From Ivison et al. (1998) \\
\end{flushleft}
\end{table*}

We now describe the observed properties of the emission lines of image
B and C.  The velocity width (FWHM) corrected for instrumental
resolution and the relative flux values measured by Gaussian
fitting procedure are listed in Table 1 and 2, respectively. The flux
values are normalized to \CIV\ lines.
 We corrected for the reddening by the Galaxy using the value of the extinction 
in this region, A$_V$=0.67 and A$_I$=0.36, derived by Ben\'itez et al. (1999).
No correction for the internal reddening
was applied since the rest-frame wavelength of the emission lines concerned here
are rather close. We treat the unresolved doublet lines as a single
line.

\begin{table*}
\caption{Relative flux of the emission lines $^{a}$}
\begin{flushleft}
\begin{tabular}{lccccccc}
\noalign{\smallskip}
\hline
\noalign{\smallskip}
 & MG2016+112 B & MG2016+112 C & F10214$^{b}$ & SMM02399$^{c}$ & HzPRG$^{d}$&
 Quasar BLR$^{d}$& Seyfert 2$^{d}$ \\
\noalign{\smallskip}
\hline
\noalign{\smallskip}
  Ly$\alpha$      & 2.73 & 3.81 & 0.53 & 3.39 & 2.17 & 8.52 & 4.52 \\
  \NV\            & 0.66 & 1.19 & 0.74 & 2.20 & 0.57 & 0.42 &      \\
  \CIV\           & 1.00 & 1.00 & 1.00 & 1.00 & 1.00 & 1.00 & 1.00 \\
  \HeII\          & 0.28 & 0.55 & 0.53 & 0.36 & 0.12 & 0.87 & 0.17 \\
  \CIII\          & 0.40 & 1.32 &      & 2.70 & 0.45 & 0.49 & 0.46 \\
  \CIII\ narrow   & 0.17 & 0.51 & 0.34 &      &      &      &      \\
  \CIII\ broad    & 0.29 & 1.25 & 0.28 &      &      &      &      \\
\noalign{\smallskip}
\hline
\noalign{\smallskip}
\end{tabular}
\\
(a) Normalized to \CIV\ line. \\
(b) From Searjeant et al. (1998) \\
(c) Evaluated by us from the values of equivalent width in 
Ivison et al. (1998) \\
(d) From McCarthy et al. (1993) \\
\end{flushleft} 
\end{table*}

\begin{figure} 
\psfig{file=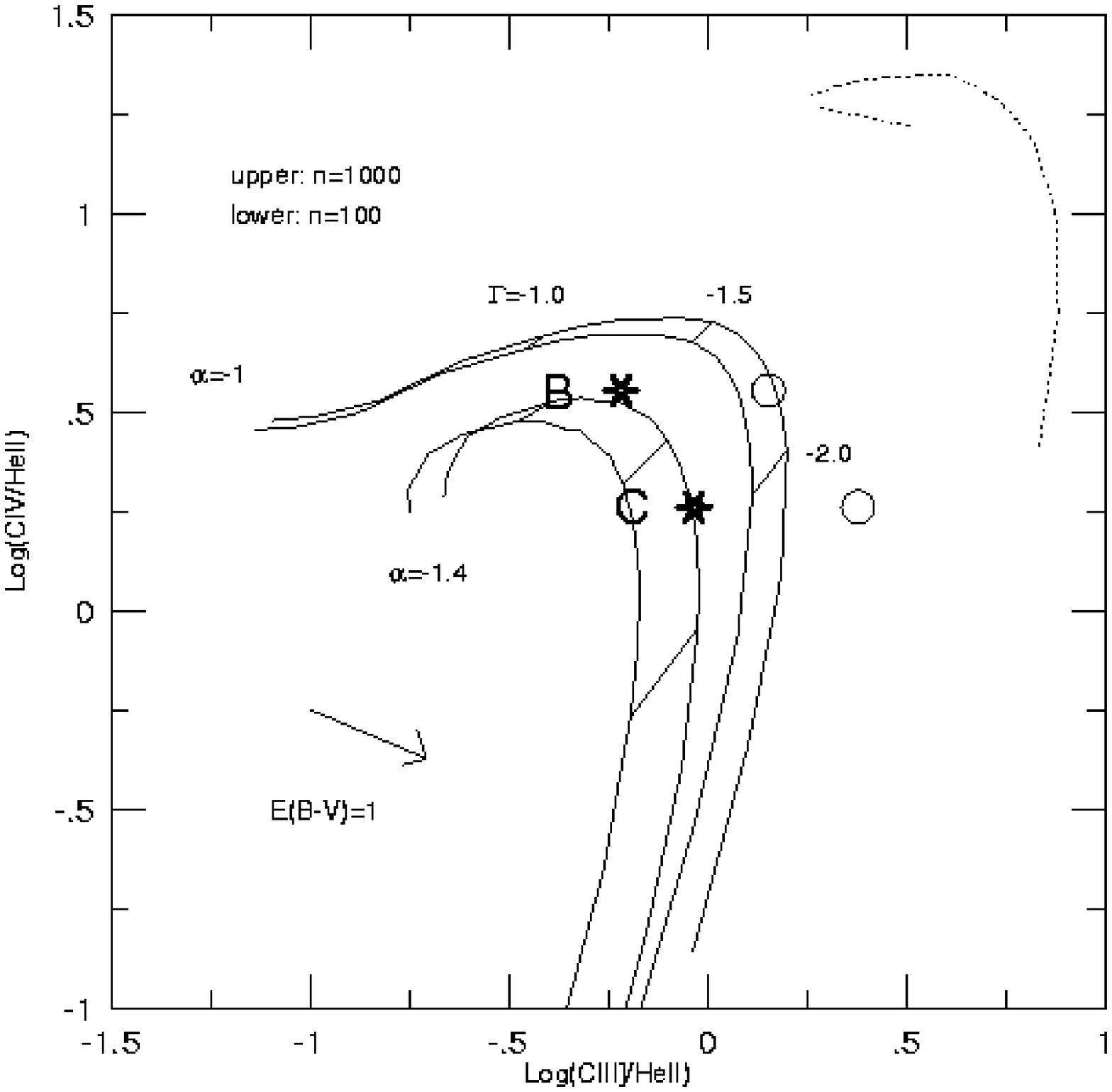,width=0.4\textwidth}
\caption[fig4.ps]{Flux
  ratios of the emission lines seen in the spectrum of image B and
  C. Asterisks show those using only the narrow component of the
  \CIII\ line and circles those using the \CIII\ flux obtained with
  single-component fitting. The grids show the line flux ratio
  predicted by photo-ionization models calculated using
  CLOUDY90 (Ferland et al. 1988). The cases of hydrogen density of 100
  cm$^{-3}$ and 1000 cm$^{-3}$ and  input power-law spectra with
  energy index $-1.0$ and $-1.4$ are plotted for ranges of
  ionization parameter log$\Gamma$ = $-1.0$ to $-2.5$ (solid lines). We
  also show the case of very high hydrogen density ( dotted line,
b$_{\rm H}$ = $10^8$ cm$^{-3}$, $\alpha=-1$). The arrow shows the 
effect of the reddening calculated with the Calzetti's relation for starburst galaxies.4
and $E(B-V)=1$. }  
\end{figure}

The velocity width of the lines are between 450-900 \kiro . These
lines are much narrower than quasar broad lines which have typical
width of 5000-10000 \kiro\ but as narrow as those of HzPRG, 500-1000
km s$^{-1}$ (e.g., McCarthy 1993). In tabel 2, we compare the
emission-line width to the two infrared-selected
gravitationally-lensed type-2 AGN at high redshift, namely \F10214\
(e.g., Rowan-Robinson et al. 1991) and \SMM\ (Ivison et al. 1998).

\CIII\ line of both B and C seem to have broad wings. Figure 3 shows
the results of two-component Gaussian fitting of the \CIII\ lines
for B and C. They cannot be perfectly fitted with a single component
but can be better fitted with a narrow plus broad components. It
seems strange that the \CIII\ line has a broad component while the
\CIV\ line, which comes from the gas at higher ionization stage and
may be closer to the central engine, shows only a narrow one. It may be
due to the extinction of the broad component of the \CIV\ line.
Another possibility is that the wings are the lines of other ions
contaminating this spectral range .

The broad \CIII\ feature is not a unique characterestics of MG2016+112
lensed object. \F10214\ (Searjeant et al. 1997) and \SMM\ (Ivison et
al. 1998) also show similar property. Searjeant et al. (1997) argue the
possibility of contamination by SiIII lines for the blue-wing
feature of the \CIII\ line in the spectrum of \F10214\ but the model
cannot fully explain the observed feature. It is interesting that
those two gravitationally-lensed type-2 AGN as well as MG2016+112
 show evidence of a broad \CIII\ line.

Next, we investigate the line flux ratios. There are significant
differences between line ratios of component B and C. While
the \HeII /\CIV\ ratio of image B is 0.28, the ratio of C is 0.51.
Also, the \CIII /\CIV\ ratio of B and C is 0.17 and 0.51, respectively
if we consider the narrow components. If we fit the lines with single
Gaussian component, then the ratio is 0.40 and 1.32.  The emission-line
gas in image B seems to be at higher ionization stage than in
C. In Fig. 4, we compare these line ratios with those predicted by
photoionization models (CLOUDY90, Ferland 1988).
We examined the cases with hydrogen density n$_{\rm H}$=100 and 1000
cm$^{-3}$, power-low ionization continuum with  $\alpha$=1
and 1.4, and a range of ionization parameter,
$\Gamma$=$10^{-2.5}$-$10^{-1}$ and solar-abundance. The obtained line
ratios are fairly consistent with the typical photo-ionization models
considered for narrow-line region of AGN.
 The differences in line ratios between B and C may be
interpreted as the difference in ionization parameters.

There may be an effect of the reddening by dust in the object on the observed line
ratio. According to Calzetti's reddening fomula for starburst
galaxies (Sawicki \& Yee 1998), the reddening of $E(B-V) = 1$ shifts
the observed values with $\Delta$log(\CIV /\HeII )=$-0.12$ and 
$\Delta$log(\CIII /\HeII ) =$+0.30$, respectively, 
which is shown by the arrow in Fig. 4. Even if
there is a fairly large amount of reddening, $E(B-V) \sim 1$, the
discussion we give here is not so much affected.

There are weak but fairly significant \NV\ lines in the spectra of B
and C. In Fig. 5 we plotted the observed \NV\ line ratios as well as
those predicted by photoionization models. Clearly, photoionizatin
models that can explain the line ratios at lower ionization state are
not consistent with the observed \NV\ flux. The observed \NV\ lines
are more than several times stronger than the predicted
value. Reddening correction moves the points further away from the
models. The origin of \NV\ line may be different from those of other
lines since the ionization potential of N$^{++++}$ is 77.5 eV which is
significantly higher than those of \CIV\ (47.9 eV), \CIII\ (24.4 eV),
and \HeII\ (24.6 eV). It may also be due to the nitrogen over
abundance. In fact, the situation is not special for MG2016+112 B and
Hamman \& Ferland (1993) also compare their photoionization models
with the observed lines of broad-line region (BLR) of quasars and
found a large nitrogen over abundance. There is also similar
nitrogen problem for the optical \NII\ lines observed in the spectra
of local AGN (e.g., Osterbrook 1986).

The fast shock models (Mouri \& Taniguchi, private communication) can
provide the \NV /\CIII\ and \NV /\HeII\ ratios that match the observed
values but then predict too strong \CIV\ lines. The line ratios
observed in image B and C are not perfectly understood with simple
photoionization or shock models.

In Fig. 6, we also plot typical line ratios of various types of
high-redshift AGN taken from McCarthy (1993), ultra-steep spectrum
HzPRGs compiled by R\"oettgering et al. (1997), and the 
infrared-selected type-2 quasars, \F10214\ (Serjeant et al. 1997).
Many of the HzPRGs in R\"oettgering et al. (1997) are distributed at
around log(\CIV /\HeII ) $\sim 0.2$ and log(\CIII /\HeII ) $\sim
-0.2$, which is consistent with the average flux ratio given by
McCarthy (1993). The line ratios of image B is very different
from typical HzPRGs.  The line ratios of C and \F10214\  lies
near the edge of the distribution of line ratios of HzPRGs.

We also plotted the line ratio of the {\it broad lines} observed in
quasar spectra compiled by Baldwin et al. (1979). They are very
different from both HzPRGs and image B and C. For a comparison, we
show the predictions of photoionization models with very
high-density gas with n$_{\rm H}$=$10^8$ cm$^{-3}$ in Fig. 4
considering that the density may be as high as 10$^{7-9}$ cm$^{-3}$ in
the quasar BLR.  The line ratios of {\it broad lines} quasars in
Fig. 6 are distributed near this model prediction.

\begin{figure} 
\psfig{file=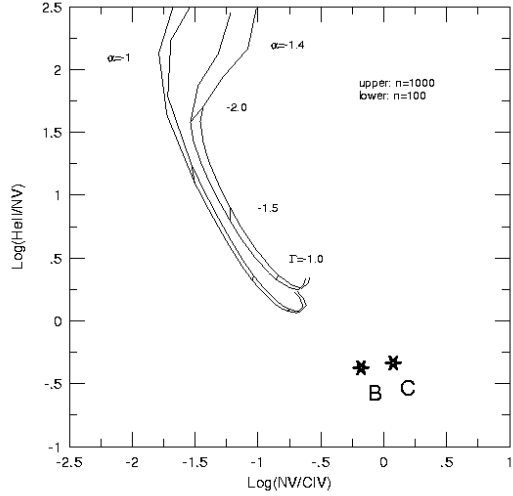,width=0.4\textwidth}
\caption[fig5.ps]{Same as Fig. 4, but for \NV\ line.}  
\end{figure}

\begin{figure} 
\psfig{file=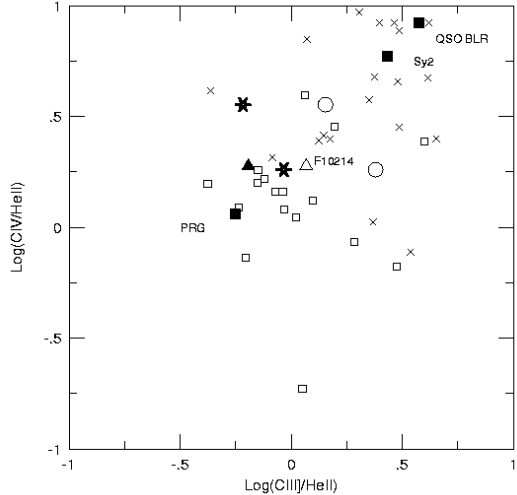,width=0.4\textwidth}
\caption[fig6.ps]{Line ratios for various kinds
  of high-redshift AGN. The solid squares are the average line ratios
  given in McCarthey (1993) and the open squares are those of USS
  HzPRGs in R\"ottgering et al. (1997). Triangles are the
  infrared-selected type-2 AGN, \F10214 , using the narrow-component
  (filled) and total (open) \CIII\ flux. The ratios of broad lines
  quasar quoted from Baldwin (1979) are shown as crosses.}
\end{figure} 

\section{Discussion}

\subsection{Lens Models}

Several lens models have been proposed for MG2016+112 
(Narasimha et al. 1984, 1987, 1989; Langston et al. 1991; Nair \&
Garrett 1997; Ben\'\i tez et al. 1999). These models are constructed to
match the optical and/or radio  data such as image
positions and flux ratios of the various components. The results
of our spectroscopic observations can be used to put further
constraints on the lens model.

As discussed in Sect. 3, the difference of \CIII , \CIV , and
\HeII\ line ratios in image B and C can be interpreted as a
difference in ionization degree. Photoionization model predict smaller
ionization parameter for C, which is a natural consequence if the
component C is dominated by the light of a region at $\sim$1.5-2 times
larger radius from the nucleus than the source region of the component
B (assuming similar density).
It is unlikely that the differences in line  ratios of 
B and C is due to the contamination by a possible radio galaxy at the
similar redshift assumed as the counterpart of the bright
flat-spectrum radio component C$_1$. If the high \CIII /\CIV\ ratio
observed in the spectrum of image C is due to such a contamination,
for example, the assumed radio galaxy should have \CIII /\CIV\ ratio
larger than $\sim 1$. Radio galaxies rarely show such high \CIII
/\CIV\ ratio.

Our spectroscopic results thus support lens models like the ones proposed by
Langston et al. (1991) and Ben\'\i tez et al. (1999). In these models,
the AGN and the narrow-emission-line region is located somewhat outside
the fold caustic in the source plane, which results in forming the
images  A and B. The outer regions of the source extends over
the caustic and the region which is very close to the caustic is
largely amplified and form an arc-like lensed images at the
position of component C. Radio sources at position C are
interpreted as a lensed image of a 'jet'  which extends
inside the caustic. The counter images of the component C will exist at
position  A and B but is mostly swamped by the AGN light due to a smaller
amplification factor (compared to position C).

\subsection{The Highest-Redshift Type-2 Quasar ?}

Since only the narrow emission lines are observed, the object is
likely to be an obscured AGN. Obscured AGN may be classified to either
of powerful radio galaxies or radio-quiet type-2 quasars if the
intrinsic power of the AGN is so powerful to be regarded as a quasar.

The number of known radio-quiet type-2 quasars (hereafter simply
referred as type-2 quasars) are still very small, which may be due to
selection effects. Previous quasar surveys with methods such as UV
excess, objective prism, soft X-ray are not sensitive in searching for
type-2 quasars. There are only several examples of type-2 quasars at
high redshift which are serendipitously discovered in far-infrared,
sub-mm, and X-ray source surveys. If MG2016+112 at z=3.27 is a
type-2 quasar, it would be the highest-redshift type-2 quasar and thus
provides a unique and important example for further studies of type-2
quasars.

\subsubsection{Intrinsic Power of MG2016+112}

The observed radio flux density of image B at 1.47 GHz is 61.7
mJy (Lawrence et al. 1984), which corresponds to a luminosity
density $L_{\rm 2.7 GHz} = 2.3 \times 10^{34}$ erg s$^{-1}$ Hz$^{-1}$
$A_{\rm GL}^{-1}$ and $L_{\rm 8.4 GHz} = 9.1 \times 10^{33}$ erg
s$^{-1}$ Hz$^{-1}$ $A_{\rm GL}^{-1}$ at 2.7 and 8.4 GHz, respectively,
with a luminosity distance of 26.5 Gpc and a spectral index
$\alpha=-0.81$ (Lawrence et al. 1984). $A_{\rm GL}$ is a
gravitational lensing amplification factor. According to the recent
lens model by Beni\'\i tez et al. (1999), the amplification factor of
image B is estimated to be $\sim 6$.

The radio power of MG2016+112 is fairly large even corrected from the
lensing amplification (e.g., Lawrence et al. 1984). Danese et
al. (1987) obtained radio luminosity functions of at 2.4 GHz for
various type of objects at z$\sim 0$. Even the most luminous local
objects have radio power $\sim 10^{32}$ erg s$^{-1}$ Hz$^{-1}$, which
is more than a order of magnitude fainter than MG2016+112. Dunlop and
Peacock (1990) evaluated radio luminosity function of the
radio-selected quasars and radio galaxies. Image B has radio
power as large as those of very luminous radio sources at z$\sim
0.5$ which are definitely categorized as `quasars' or `powerful radio
galaxies' ($10^{33}-10^{34}$ erg s$^{-1}$ Hz$^{-1}$ at 2.7 GHz).

Bischef \& Becker (1998) recently investigated radio emissions for a
sample of 4079 known quasars based on the NVSS radio catalog. These
quasars are selected from the Ver\'on-Cetty \& Ver\'on (VCV) catalog
and constitute the largest compilation so far to study the radio
properties of quasars based on homogeneous radio observations in one
frequency. In their Fig. 4, they presented a distribution of radio
power at 8.4 GHz with redshift. The bimodal distribution of the radio
power is evident at least to z$\sim 2.5$. At higher redshift, the
number of known radio-luminous quasars is too small to draw firm
conclusion, but the tendency seems to hold.  The radio power of
MG2016+112 at 8.4 GHz is likely to lie on the extension of the
lower sequence, if we adopt an amplification factor of $\sim 10$. In
Fig. 7 we plot the observed and amplification-corrected radio power of
MG2016+112 B superposed on the figure of Bischef \& Becker's
(their Fig. 4).

\subsubsection{Radio Loudness}

Radio loudness of AGN is conventionally defined by the radio to
optical (or ultra violet) flux ratio. Bischef \& Becker (1998) also
obtained the distribution of the radio loudness for VCV quasars.
According to their definition, radio-loud quasars have 
log $(L_{\rm 8.4 GHz}/L_B)$ larger than 1. The upper sequence in Fig. 4 of
Bischef \& Becker (1998) corresponds to radio-loud
quasars and the lower to radio-quiet quasars.  It is difficult,
however, to evaluate the radio loudness of MG2016+112 by using
the radio to optical flux ratio since it is an obscured narrow-line
object. We observe (at most) only a scattered light of the nucleus in
the optical wavelength and the contamination by light from the host
galaxy must be relatively large. The nominal value, log $L_{\rm 8.4
  GHz}$/$L_B$ = 3.4, does not mean that this object is a
radio-loud AGN. Indeed, as mentioned in the previous subsection, the
observed radio power of MG2016+112 after lensing amplification
correction seems too faint to be classified as radio-loud,
which suggests that this object is a {\it radio-quiet} AGN. It is not
surprising that the scattered component is more than hundred times
smaller than the intrinsic luminosity and the intrinsic log $L_{\rm
  8.4 GHz}$/$L_B$ value of the object can be $\la 1$.

\begin{figure} 
\psfig{file=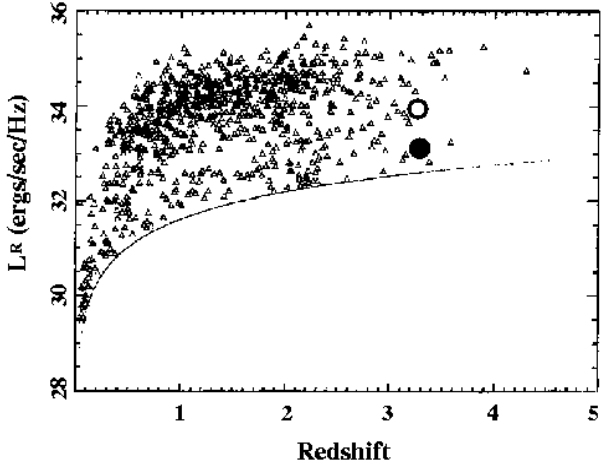,width=0.5\textwidth}
\caption[fig7.ps]{The observed (thick open circle) and the
  lensing amplification corrected (filled circle) radio luminosities of
  MG2016+112 superposed on  Fig. 4 of Bischof \& Becker
  (1990) (scanned by us) which shows the distribution of radio
  luminosity at 8.4 GHz vs redshift for VCV quasars. Crosses are
  values for the broad lines of type-1 quasars.}  
\end{figure}

How typical the radio power is if MG2016+112 is a radio-quiet
AGN ? Figure 7 suggests that it may be one of the most radio active
object among the radio-quiet quasars at this redshift.  Kukula et al.
(1998) recently investigated the correlation between radio and optical
luminosity of a sample of nearby radio-quiet quasars and Seyfert
galaxies (z $< 0.2$). The most luminous nearby quasars with M$_V \sim
-26$ have log $L_{\rm 8.4 GHz}$ $\sim 10^{31}$ erg s$^{-1}$ Hz$^{-1}$.
If the correlation holds for high-redshift quasars which are typically
50-100 times more luminous than nearby AGNs, MG2016+112
(L$_{\rm 8.4 GHz}$ $\sim 10^{33}$ erg s$^{-1}$ Hz$^{-1}$) may be one
of the most luminous quasars in optical wavelength, too, even if there is
some radio excess.

\subsubsection{Morphologies}

HST NICMOS observations (Falco et al.) revealed that the rest-frame
optical light of MG2016+112 A and B is dominated by the point sources.
The resolution limit of NICMOS is $\sim 0.15$ arcsec. If we adopt the
lens model of Ben\'\i tez et al. (1999), the size of
the unresolved lensed object must be smaller than 0.03 arcsec, which
corresponds to $\sim 200$ pc at z=3.27. It may not be surprising that
A and B are not resolved if we observe only the obscured nucleus or its
scattered light which may come from inside the narrow-line region.

Many of HzPRGs show resolved faint extension in NIR images
which may be star lights of their host galaxies. On the other hand,
the known type-2 quasars at z$\sim 1$-2.5 are not resolved or only
marginally resolved (Ohta et al. 1995; Almani et al. 1995; Ivison et
al. 1998). There could be some difference between observable
properties of host galaxies of HzPRGs and type-2 quasars.
 
The radio images of A and B are also point-like with $\sim 15$ mas
resolution (Garrett et al. 1996). 15 mas corresponds to about 40 pc in
the physical scale at z=3.27.  Kukula et al. (1998) presented the 1.4
GHz maps of the nearby radio-quiet quasars with $\sim 0.5$ arcsec
resolution. They found a significant fraction of the radio emission in
radio-quiet quasars originates in a compact nuclear source directly
associated with the quasar. Therefore the point-like morphology of the
radio emission of MG2016+112 is not surprising if it is a
radio-quiet AGN. At the same time, some outer structures are also seen
in the radio maps of the nearby radio-quiet quasars. The maximum
extent of the radio emission of resolved sources is typically a
few kpc. It is possible that such a structure associated with the
lensed object extends over the diamond caustic of the source plane
to form the strongly amplified radio image at the position of 
component C.

\subsubsection{Summary of the Type-2 Quasar Nature of MG2016+112}

The central engine of MG2016+112 must be obscured since only the
narrow emission lines are observed. The radio and optical morphologies
are both compact, which is not compatible with typical properties of
HzPRGs but common for radio-quiet AGNs. The radio power is much
smaller than typical radio-loud quasars at z$\sim 3$ but consistent
with luminous radio-quiet quasar.  Finally, the emission line flux
ratios of image B are not compatible with typical HzPRGs. We thus
conclude that MG2016+112 is not a typical powerful radio galaxies but
naturaly classifies as a rare example of a high-redshift radio-quiet
type-2 quasar.

Since \F10214\ and \SMM\ are detected in CO in sub-millimeter
(Rowan-Robinson et al. 1993; Ivison et al. 1998; Frayer et al. 1998),
it will be interesting to investigate the cold gas and dust properties
of MG2016+112. As expected from the obscuration of the nucleus, large
dust content may be a general property of type-2 quasars.


\begin{acknowledgements}
 We thank Drs. H. Mouri and Y. Taniguchi for kindly providing us 
the results of the shock-model calculations. This research was 
partially supported by grants-in-aid for scientific research of the 
Japanese Ministry of Education, Science, Sports and Culture (09740168, 07055044).
JPK thanks CNRS for support and the Yamada Science Fundation for fruitful
visits in Japan. MH also thanks the financial supports of Yamada Science Foundation.

\end{acknowledgements}

\end{document}